\documentclass[12pt]{article}
\usepackage{graphicx}
\usepackage{amssymb,amsmath}
\newcommand{\llg}{\ensuremath{\ell_j \rightarrow \ell_i\gamma}}
\newcommand{\brmeg}{BR(\ensuremath{\mu\rightarrow e \gamma})}
\newcommand{\meg}{\ensuremath{\mu\rightarrow e \gamma}}
\newcommand{\mue}{\ensuremath{\mu\mbox{-}e}}
\newcommand{\tmg}{\ensuremath{\tau \rightarrow \mu\gamma}}
\newcommand{\mee}{\ensuremath{\mu\rightarrow 3e}}
\newcommand{\tb}{\ensuremath{\tan\beta}}
\newcommand{\rate}{\ensuremath{R(\mu \,Ti\rightarrow e \,Ti)}}

\begin{document}
\centerline{\bf  Constraining msugra parameters with $\boldsymbol{\meg}$ and $\boldsymbol{\mue}$ conversion in nuclei}
\vspace{1cm}
\centerline{Carlos E. Yaguna}
\centerline{\it \small Department of Physics and Astronomy, UCLA, Los Angeles, CA 90095-1547}
%\\ E-mail: \email{yaguna@physics.ucla.edu}}
\begin{abstract}We show that, in the MSSM with msugra boundary conditions and seesaw induced neutrino masses, the values of \brmeg~ and the \mue~ conversion rate in a nucleus constrain msugra parameters in a model independent way
%\preprint{UCLA/05/TEP/5}

\end{abstract}

%\maketitle
\section{Motivation}
The first evidence of supersymmetry to be found may not be the direct production of superpartners but the observation of lepton flavor violating processes. It is therefore crucial to learn how to extract supersymmetric parameters from the measured rates of such processes. In this paper we show that such a project is, at least in principle, feasible. Even if supersymmetric particles have not been detected, the observation of LFV processes can constrain the supersymmetric parameter space.

Being so large, the parameter space of the general MSSM is very difficult to handle. That is the reason why it is common to focus attention on the minimal supergravity model -msugra-, in which all the soft supersymmetry breaking scalar masses, the three gaugino masses and all trilinear couplings are required to have a common value  at the unification scale --respectively $m_0$, $M_{1/2}$ and $A_0$. By solving the renormalization group equations with these boundary conditions, the weak scale superparticle spectrum is obtained. The requirement of  radiative electroweak symmetry breaking allows the determination of the supersymmetric Higgs mixing parameter $\mu^2$ and the expression of the soft susy breaking bilinear term $B$ in terms of $\tan \beta$, the ratio of the vev's of the two Higgs fields. Hence, in msugra all sparticle masses and couplings are obtained in terms of the parameter set
\begin{equation}
m_0\,,M_{1/2}\,,A_0\,,\tan\beta\,,  \mathrm{sign}(\mu)\,.
\end{equation} 
Though msugra is not the only possible supersymmetric framework, it undoubtedly is an appealing and predictive model suitable to test new ideas and we will limit our following discussion to it.

To take into account neutrino masses, the MSSM must be extended. The simplest way of generating non-zero neutrino masses is with the addition of  right-handed neutrinos through the see saw mechanism. A remarkable prediction of this extension of the MSSM is that, through renormalization, the left-handed slepton mass matrix, $m_L^2$, acquires flavor violating (off-diagonal) contributions,
\begin{equation}
(m_L^2)_{ij}\neq 0 \quad (i\neq j)\,.
\label{off}
\end{equation}
These corrections manifest themselves at low energies by inducing lepton flavor violating processes mediated by supersymmetric particles \cite{Borzumati:1986qx}.

If those contributions were known we could predict the rate of LFV processes for any given set of msugra parameters. Unfortunately, $(m_L^2)_{ij}$ not only is not known but can not be reconstructed from low energy data. The usual way of dealing with this problem has been to assume additional flavor symmetries that reduce the number of free parameters and allow to compute such off-diagonal elements. Hence, a given symmetry typically implies a prediction for LFV processes and from it constraints on the msugra parameter space can be derived. The many studies of this kind that have been done \cite{Sato:2000ff,Kageyama:2001tn,Ellis:1999uq}, though necessarily model dependent, have shown that LFV processes are likely to be observed in planned and ongoing experiments. Moreover, due to the significant improvements on the experimental limits that they will reach \cite{Aoki:2003zh}, it may well be that supersymmetry is discovered through lepton flavor violation even before its direct detection at the LHC \cite{Masiero:2004vk}. This is, after all,  a common pattern in high-energy physics: new particles are first detected through their virtual effects and only afterward are directly produced. And such effects usually allow to constrain the particles properties or the kind of new physics that is behind them. Will the observation of LFV processes reveals us anything about supersymmetry? We shall show that the values of $\brmeg$ and $\mue$ conversion in nuclei determine the sign of $\mu$ and constrain $\tb$.

\section{Analysis}
If the supersymmetric spectrum is not known, the observation of a single lepton flavor violating process, though of great importance, cannot tell us anything about the parameter space of msugra. Indeed, different combinations of msugra parameters and off-diagonal elements  may yield the same rate for a given process. To constrain msugra in a model independent way, therefore, it is necessary to observe at least two of these processes  and to look for possible correlations among them.

Because it provides the strongest constraint on $\mue$ flavor violation, we will use the decay $\meg$ as the reference process to correlate with. Other LFV processes that might be related to $\meg$ include $\tmg$, $\mee$, and $\mue$ conversion in nuclei.

The off-shell amplitude for $\llg ^*$, that gives a significant contribution to a wide class of lepton flavor violating  processes, may be  written as
\begin{equation}
T=e\epsilon^{\alpha *}\bar u_i(p-q)\left[q^2 \gamma_\alpha (A_1^LP_L+A_1^RP_R)+m_j i\sigma_{\alpha\beta}q^\beta (A_2^LP_L+A_2^R P_R)\right]u_j(p)
\label{T}
\end{equation}
where $q$ is the momentum of the photon, $e$ is the electric charge, $\epsilon^*$ the photon polarization vector, $u_i$ and $u_j$ the wave functions of the initial and final leptons,  $p$ is the momentum of the particle $\ell_j$, and explicit expressions for the coefficients can be found  in \cite{Hisano:1995cp}. Since each coefficient receives contributions from  both chargino-sneutrino and  neutralino-slepton loops, the amplitude depends on several supersymmetric masses. On the other hand, in the mass insertion approximation is easily seen  that all four coefficients  are  proportional to the corresponding off-diagonal element and consequently $T\propto (m_L^2)_{ij}$.

The branching ratio of the decay $\meg$ is obtained from (\ref{T}), 
\begin{equation}
\brmeg=\frac{48 \pi^3 \alpha}{G_F^2} \left(\left|A_2^L\right|^2+\left|A_2^R\right|^2\right).
\label{meg}
\end{equation}
It depends only on the $A_2$ coefficients, which fulfill the relation $A_2^R\gg A_2^L$. 

$\tmg$ is a decay analogous to $\meg$. They  have the same loop  structure but are induced by different off-diagonal elements in the left-handed slepton mass matrix, $(m_L^2)_{21}$ for $\meg$ and  $(m_L^2)_{32}$ for $\tmg$. Since these two quantities are in principle independent, correlations between $\tmg$ and $\meg$ can not be used to get information about the msugra parameter space. To this end, we must search for other processes induced by the same off-diagonal element. That is, other $\mue$ transitions.

The decay $\mee$ receives contributions from penguin-type diagrams and from box-type diagrams. It turns out, however, that the amplitude is dominated by a penguin-type contribution involving the same combination of $A_2^L$ and $A_2^R$ that enters into $\meg$. Indeed,
\begin{equation}
\frac{BR(\mu\rightarrow 3e)}{\brmeg}\approx \frac{\alpha}{8\pi}\left(\frac{16}{3}\log \frac{m_\mu}{2 m_e}-\frac{14}{9}\right)= 7 \times 10^{-3}\,. 
\end{equation}
Being  constant over the whole parameter space, this ratio cannot be used to constrain it.

$\mue$ conversion in nuclei also receives contributions from penguin and box diagrams, and the $\gamma$-penguin diagram dominates over a large portion of the parameter space. In that region, the rate of $\mue$ conversion in nuclei is given by
\begin{eqnarray}
R(\mu\rightarrow e)&=&\frac{\Gamma(\mu\rightarrow e)}{\Gamma_{capt}}\\
&\simeq& \frac{4\alpha^5 Z_{eff}^4 Z |F(q)|^2 m_\mu^5}{\Gamma_{capt}} \left[ |A_1^L-A_2^R|^2+|A_1^R-A_2^L|^2\right]
\label{mue}
\end{eqnarray}
where $Z$ denote the proton number, $Z_{eff}$ is the effective charge of the muon in the $1s$ state, $F(q^2)$ is the nuclear form factor, and $\Gamma_{capt}$ is the total capture rate. Without loss of generality, we will limit our following discussion on $\mue$ conversion rates to the nucleus $^{48}_{22} Ti$. Then \cite{Hisano:1995cp,Kitano:2002mt}, $Z_{eff}=17.6$, $F(q^2\simeq -m_\mu^2)\simeq 0.54$, and $\Gamma_{capt}=2.59\times 10^6 s^{-1}$.

Owing to the additional dependence of \rate~  on the $A_1^{L,R}$ amplitudes, the ratio
\begin{equation}
\frac{\brmeg}{\rate}\equiv C
\label{cc}
\end{equation}
is not expected to be constant. Its value might contain useful information about supersymmetric parameters. In fact, since both $\brmeg$ and $\rate$ are proportional to $|(m^2_L)_{21}|^2$, $C$ does not depend on it. Thus, $C$ is determined exclusively by msugra parameters.

We have thus identified $\mue$ conversion in a nucleus as a process that in conjunction with $\meg$ could allow us, through $C$, to constrain the msugra parameter space. In the following section, we will investigate numerically such constraints.

\section{Results}
Now we proceed to evaluate $C$, as defined in (\ref{cc}), in msugra models with seesaw mechanism of neutrino mass generation. The method we follow  consists of scanning randomly the relevant set of soft-breaking terms. For each set we evaluate $C$  and display the results for different sets  as scatter plots. In such plots, correlations, if they exist, must be evident.

We will explore the following ranges for msugra parameters
\begin{eqnarray}
5\leq &\tb & \leq 50\,,\\
200 ~\mbox{GeV}\leq &m_0& \leq 1000~\mbox{GeV}\,,\\
200 ~\mbox{GeV}\leq &M_{1/2}& \leq 1000~\mbox{GeV}\,,
\label{msugra}
\end{eqnarray}
setting $A_0=0$\footnote{This is just a simplifying assumption which does not affect our results.} and taking into account both possible signs of $\mu$ ($\pm$).

Concerning the origin of lepton flavor violation we will only assume that the dominant contribution to lepton flavor violating processes resides in the left-handed slepton mass matrix, as it happens in seesaw models, but without considering any particular structure for the neutrino Yukawa couplings or any specific value for the off-diagonal elements. Instead, we parametrize the matrix element $(m^2_L)_{21}$ -the relevant one for $\mue$ transitions- at low energies as
\begin{equation}
(m^2_L)_{21}=(m^2_L)_{22}\times 10^{\rho}\,,
\end{equation}
being $\rho$ a free parameter. A suitable range for it is
\begin{equation}
-1\leq  \rho \leq -5\,.
\label{rho}
\end{equation}
This low energy parametrization guarantees that our results are valid not only in see saw models but also in all models where lepton flavor violation is induced by the left-handed slepton mass matrix.

In the ranges we are considering for msugra parameters and the off-diagonal element, $\brmeg$ and $\rate$  vary over 10 orders of magnitude, confirming that it is not possible to make model independent predictions for these processes.  A large part of such variation, however, comes from the off-diagonal element and cancels out in $C$.

\begin{figure}
\includegraphics[angle=-90,scale=0.7]{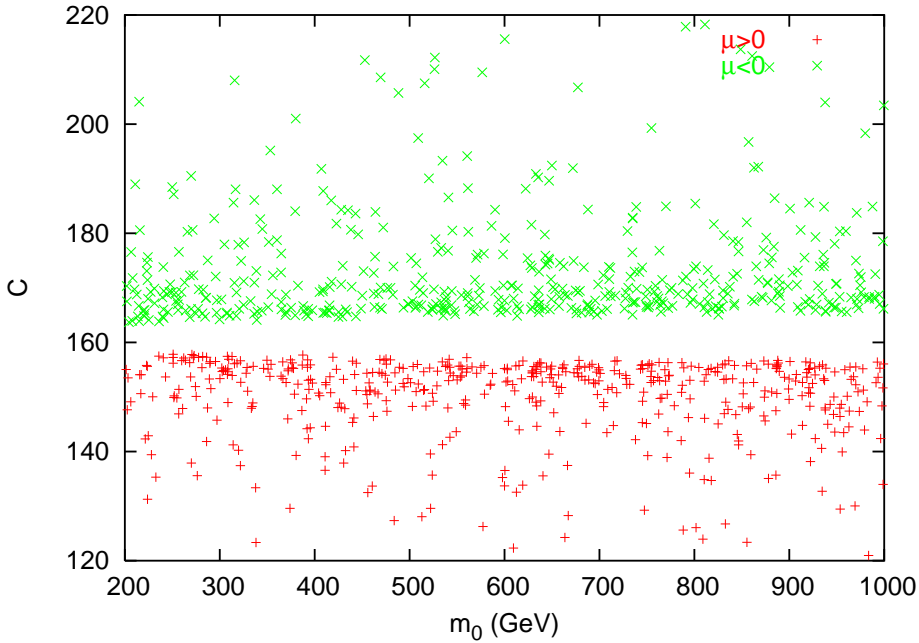} 
\caption{Scatter plot of $C$ as a function of $m_0$ for $\mu>0$ and $\mu<0$.}
\label{scatm0}
\end{figure}

\begin{figure}
\includegraphics[angle=-90,scale=0.7]{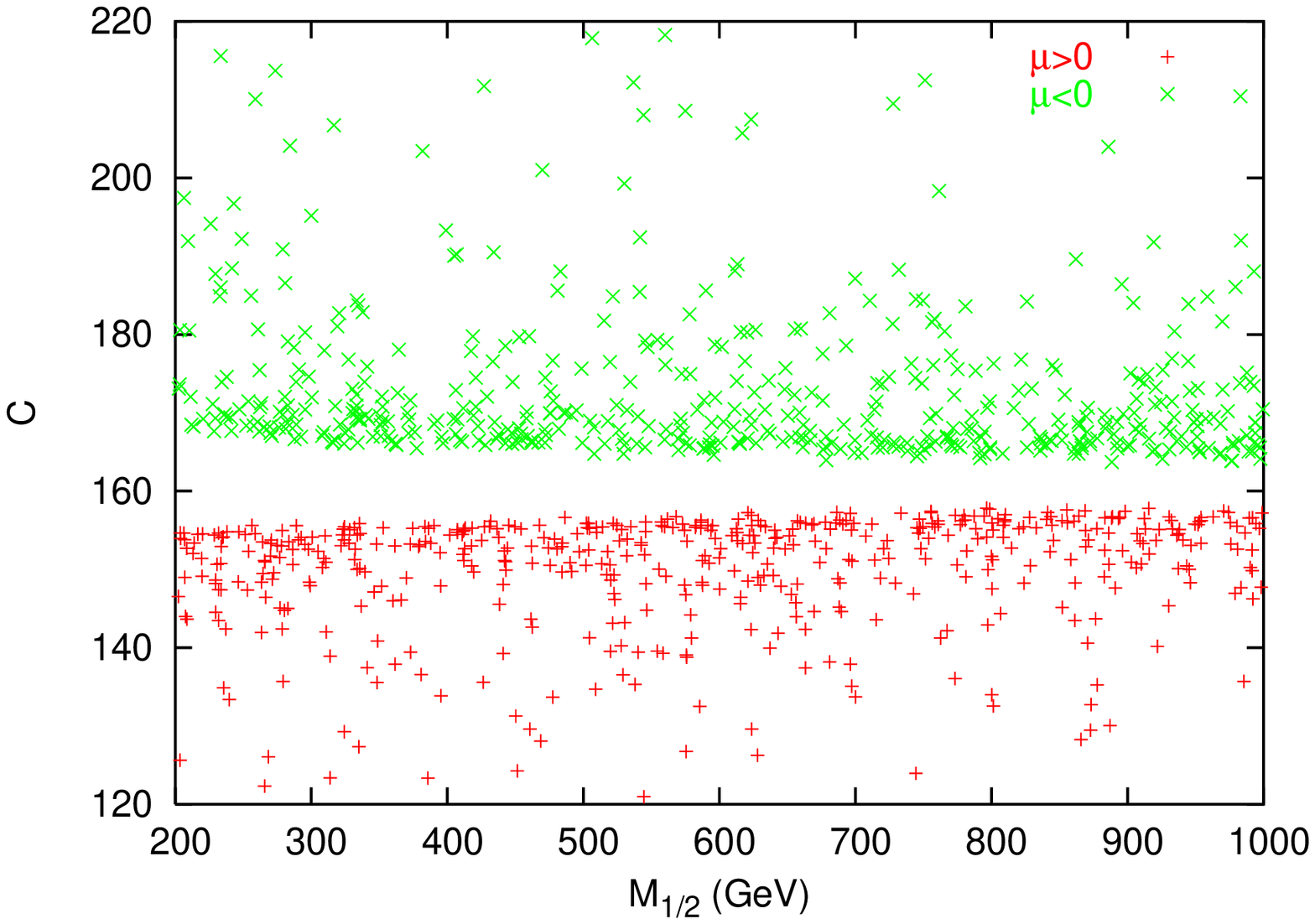}
\caption{Scatter plot of $C$ as a function of $M_{1/2}$ for $\mu>0$ and $\mu<0$.}
\label{scatm12}
\end{figure}
Scatter plots of $C$ as a function of $m_0$ and $M_{1/2}$ are shown in figures \ref{scatm0} and \ref{scatm12}. The points are concentrated in the narrow interval $130\leq C \leq 190$ and there is a gap between points with $\mu> 0$ and those with $\mu< 0$. These features can be easily understood. The coefficients in (\ref{T}) satisfy
\begin{equation}
\left|A_2^R\right|> \left|A_1^L\right|> \left|A_L^2\right|\gg \left|A_1^R\right|\label{dom}
\end{equation}
and so $\brmeg\propto |A_2^R|^2$ whereas $\rate \propto |A_2^R-A_1^L|^2$. Moreover, $\mathrm{sign}(A_2^R)=\mathrm{sign}(\mu)$ whereas $A_1^L$ is always negative. Thus, the gap observed in the figures comes from $\rate$  and is due to the change in the relative sign between $A_2^R$ and $A_1^L$ when $\mu$ changes sign. As expected from these considerations, $C(\mu<0)$ is always larger than $C(\mu>0)$, and the gap between them is small because one of the two -$|A_2^R|$- is much larger than the other.

Since practically any given $C$ is compatible with all possible values of $m_0$ and $M_{1/2}$,  $C$ cannot constrain them. $C$ does determine the sign of $\mu$, however. For instance, $C\sim 180$ implies $\mu<0$ whereas $C\sim 150$ implies $\mu>0$.

\begin{figure}
\includegraphics[angle=-90,scale=0.7]{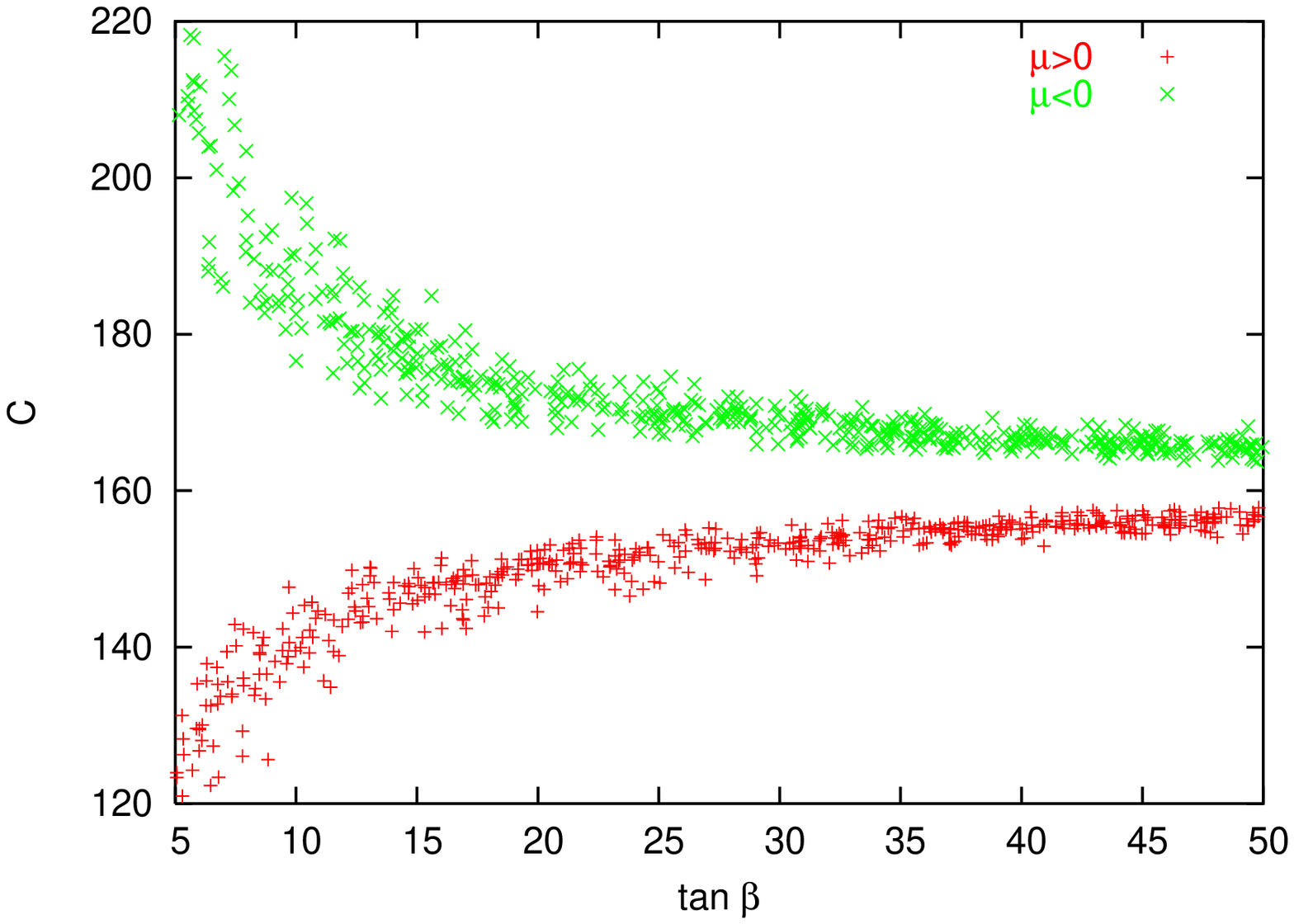}
\caption{Scatter plot of C as a function of $\tb$ for $\mu>0$ and $\mu<0$.}
\label{scat}
\end{figure}

More interesting is the correlation of $C$ with $\tb$ (see figure \ref{scat}). The coefficient $A_1^L$ is independent of $\tb$ whereas $A_2^R\propto \tb$. Hence, at low $\tb$ the interference between them in $\rate$ is stronger, giving rise to a larger gap.  When $\tb$ increases, $A_2^R$ becomes much larger  than $A_1^L$ and the gap gets consequently reduced. 

Due to the strong dependence of $C$ on $\tb$,  not all possible values of $\tb$ are compatible with a given $C$. Thus, $\tb$  can be constrained. For example, $C\geq 180$ would exclude the region $\tb \geq 20$.

Thus, we have demonstrated that  $C$ correlates nicely with the sign of $\mu$ and with $\tb$. If $C$ is measured it becomes possible to determine from it the sign of $\mu$ and to constrain $\tb$ --at least in principle. Indeed, to establish whether such measurements can be performed to the required accuracy is beyond the scope of the present work. By itself, that such neat and clear correlations between msugra parameters and lepton flavor violating observables exist independently of any particular flavor model is a new and remarkable result.

Possible extensions of this work would include the search for correlations in supersymmetric frameworks beyond msugra; the study of the uncertainties associated with the measurement of $C$ and their implications for extracting supersymmetric parameters; the inclusion of higher order (two-loop) contributions to lepton flavor violating processes; the incorporation of all cosmological and phenomenological constraints on the parameter space. All these are   interesting paths to be explored.

\section{Conclusion}
We have shown that in msugra models the values of $\brmeg$ and the $\mue$ conversion rate in a nucleus  determine the sign of $\mu$ and constrain $\tb$ practically in a model independent way. In fact, this result holds as long as the dominant source of lepton flavor violation resides in the left-handed slepton mass matrix. In particular, it is valid, independently of the value of the off-diagonal  elements, in all models with seesaw induced neutrino masses.

\end{document}